\documentclass{mykapproc} 
\usepackage{graphicx,amsfonts}

\normallatexbib

\begin{document}

\newcommand{\ket}[1]{\mbox{$|#1\rangle$}}
\def\ua{\uparrow}
\def\da{\downarrow}
\def\be{\begin{equation}}
\def\ee{\end{equation}}

\articletitle[]{Quantum Computing with Electron Spins in Quantum Dots}
\chaptitlerunninghead{Quantum Computing with Electron Spins in Quantum Dots}

\author{L.M.K. Vandersypen, R. Hanson, L.H. Willems van Beveren, J.M.\\ Elzerman, J.S. Greidanus, S. De Franceschi and L.P. Kouwenhoven}
\affil{Dept. of Applied Physics, DIMES, and ERATO Mesoscopic Correlation Project,\\ Delft University of Technology\\PO Box 5046, 2600 GA Delft, the Netherlands}
\email{lieven@qt.tn.tudelft.nl}

\begin{abstract}
We present a set of concrete and realistic ideas for the implementation of a small-scale quantum computer using electron spins in lateral GaAs/AlGaAs quantum dots. Initialization is based on leads in the quantum Hall regime with tunable spin-polarization. Read-out hinges on spin-to-charge conversion via spin-selective tunneling to or from the leads, followed by measurement of the number of electron charges on the dot via a charge detector. Single-qubit manipulation relies on a microfabricated wire located close to the quantum dot, and two-qubit interactions are controlled via the tunnel barrier connecting the respective quantum dots. Based on these ideas, we have begun a series of experiments in order to demonstrate unitary control and to measure the coherence time of individual electron spins in quantum dots.
\end{abstract}
\begin{keywords}
quantum computing, quantum dots, electron spin resonance
\end{keywords}

\section{Introduction}

The spin of a single electron placed in a static magnetic field $\vec{B_0}$ provides a natural two-level system suitable as a qubit in a quantum computer~\cite{Nielsen00b}. Loss and DiVincenzo proposed to isolate individual electrons in a quantum dot array
and showed that the electron spins can in principle be initialized, coherently manipulated and read out~\cite{Loss98a}. While there has been continued theoretical work in this area~\cite{Burkard99a,Golovach02a}, the Loss-DiVincenzo proposal has not yet been realized experimentally. We have recently begun such experiments and present here the concrete and realistic path we are taking, towards the experimental demonstration of single- and two-qubit gates and the creation of entanglement of spins in quantum dot systems.


\section{Qubit}

The qubit is represented by the spin of a single electron in a quantum dot defined by electrostatic gates on top of a GaAs$/$Al$_x$Ga$_{1-x}$As heterostructure. The spin ground state $\ket{\ua}$ and excited state $\ket{\da}$ are separated in energy by the Zeeman splitting $\Delta E_z = g_d \mu_B B_0$, with $g_d$ the dimensionless $g$-factor in the dot, $\mu_B$ the Bohr magneton and $B_0$ the static magnetic field strength. Taking $g_d = 0.44$, the bulk value in GaAs (we neglect the sign), we have $\Delta E_z \approx 25 \mu$eV per Tesla in $B_0$.

For comparison, the charging energy of a dot, $e^2/C$, with $C$ the  total capacitance of the dot, is typically a few meV, much larger than the Zeeman energy. The discrete energy level spacing is about a meV in small dots, also larger than $\Delta E_z$~\cite{Kouwenhoven97a,Kouwenhoven01a}. In what follows, we will assume that the electron always remains in the orbital ground state.

Although it may not be strictly necessary to work with few-electron dots, operation with one electron in each dot makes the experiment much more transparent. Following other groups~\cite{Ciorga00a,Sprinzak01a}, we have recently realized lateral, split-gate quantum dots with a controllable number of electrons down to $\ldots 2, 1, 0$ electrons~\cite{Elzerman02a}.


\section{Initialization}
\label{sec:initialization}

The primary goal of initialization is to place the qubit in a pure (i.e. well-known) state, say $\ket{\ua}$, as this is the desired initial state for most quantum algorithms~\cite{Nielsen00b}. The additional ability to initialize the qubit to $\ket{\da}$, or alternatively to a mixed state where the spin is probabilistically in $\ket{\ua}$ or $\ket{\da}$, would be very useful for testing whether the read-out schemes of Section~\ref{sec:readout} can distinguish $\ket{\ua}$ from $\ket{\da}$. We first present two methods for initialization to the pure state $\ket{\ua}$, and then discuss a variant for initialization to a mixed state.

Initialization to the ground state occurs naturally when we allow the electron spin to reach thermal equilibrium at high $B_0$ and low temperature $T$. Pr[$\ket{\ua}$], given by the Boltzman factor, is over $99\%$ when
\be
g_d \mu_B B_0 > 5  k_B T \;,
\label{eq:polarization}
\ee
with $k_B$ Boltzman's constant. This condition is easily satisfied at, e.g., 5 Tesla and 300 mK ($g_d=0.44$). 
Thermal equilibration on the dot is a very simple and robust initialiation approach, which can be used for any $B_0$ orientation and doesn't require spin-polarized leads. It takes about $5 T_1$ to reach equilibrium, which may be on the order of 1 ms ($T_1$ is the spin relaxation time).

\begin{figure}
\begin{center}
\raisebox{1.5cm}{(a)}
\includegraphics[width=5cm]{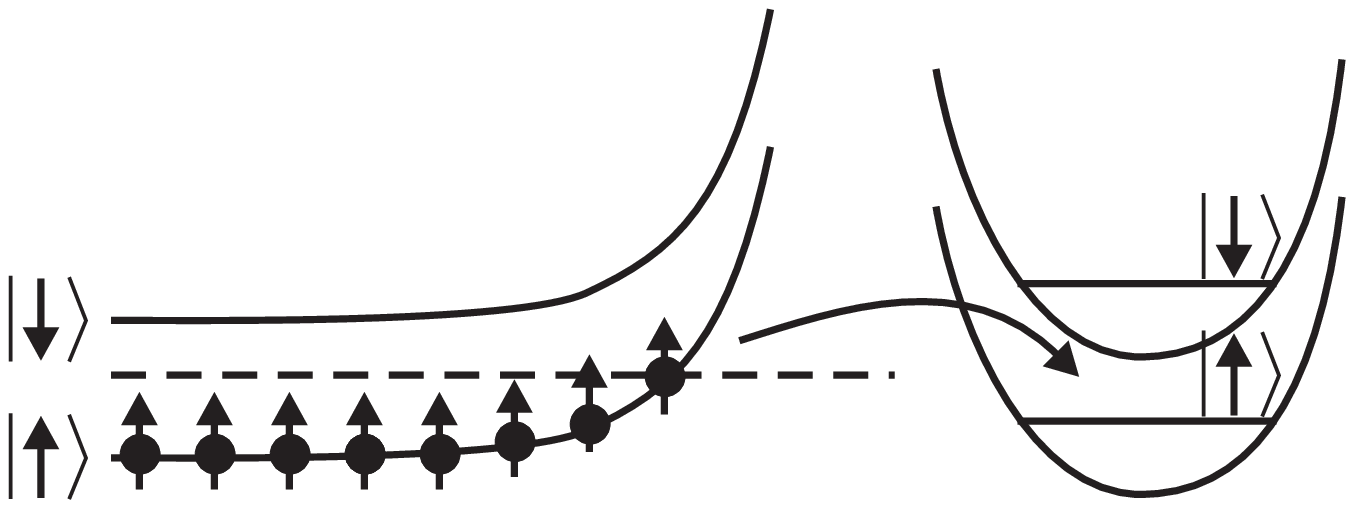}
\hspace{.5cm}
\raisebox{1.5cm}{(b)}
\includegraphics[width=5cm]{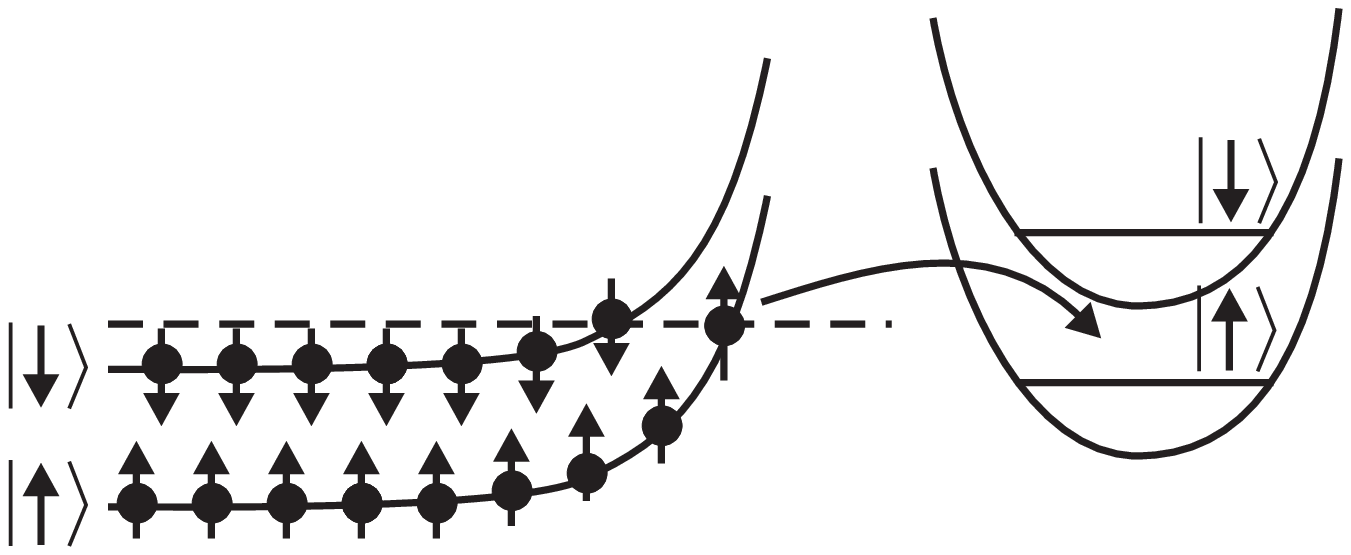}
\end{center}
\vspace*{-0.3cm}
\caption{The chemical potential for a single $\ua$ or $\da$ electron in a quantum dot, coupled to leads in the quantum Hall regime with (a) filling factor $\nu=1$ and (b) $\nu=2$. The dashed line indicates the Fermi level $E_F$ and the solid lines in the leads represent the lowest spin-split Landau level. Note that a self-consistent (in)compressible edge state picture does not affect the arguments in this paper.}
\label{fig:init_leads}
\end{figure}

Alternatively, we can let an electron tunnel to an empty dot from $\ua$-polarized leads. This gives a qubit initialized to $\ket{\ua}$, as long as the spin is conserved during tunneling, which is plausible based on transport measurements in two-dimensional electron gases (2DEGs)~\cite{Kikkawa99a}. Highly polarized leads can be obtained in the quantum Hall regime with filling factor $\nu =1$ (Fig.~\ref{fig:init_leads} a). Only the lowest spin-split Landau level is then occupied, provided $g_l \mu_B B_0 > 5  k_B T$, analogous to Eq.~\ref{eq:polarization} ($g_l$ is the $g$-factor in the leads). In fact, magnetotransport measurements in 2DEGs with odd $\nu$ have shown that for an electron in the leads to go from $\ket{\ua}$ to $\ket{\da}$, it must overcome not only the single-particle Zeeman energy but also the many-body exchange energy between the electrons in the leads~\cite{Englert82a}. We can describe this situation via an effective $g$-factor $g_{l,\mathrm{eff}}$, which can be as large as ten times $g_l$. The leads are thus spin-polarized as long as $g_{l,\mathrm{eff}} \mu_B B_0 > 5  k_B T$. Tunneling from spin-polarized leads also offers a way to robust pure-state initialization, and the tunnel time can easily be tuned under $1 \mu$s.

\begin{figure}
\begin{center}
\raisebox{3cm}{(a)}
\includegraphics[width=4cm]{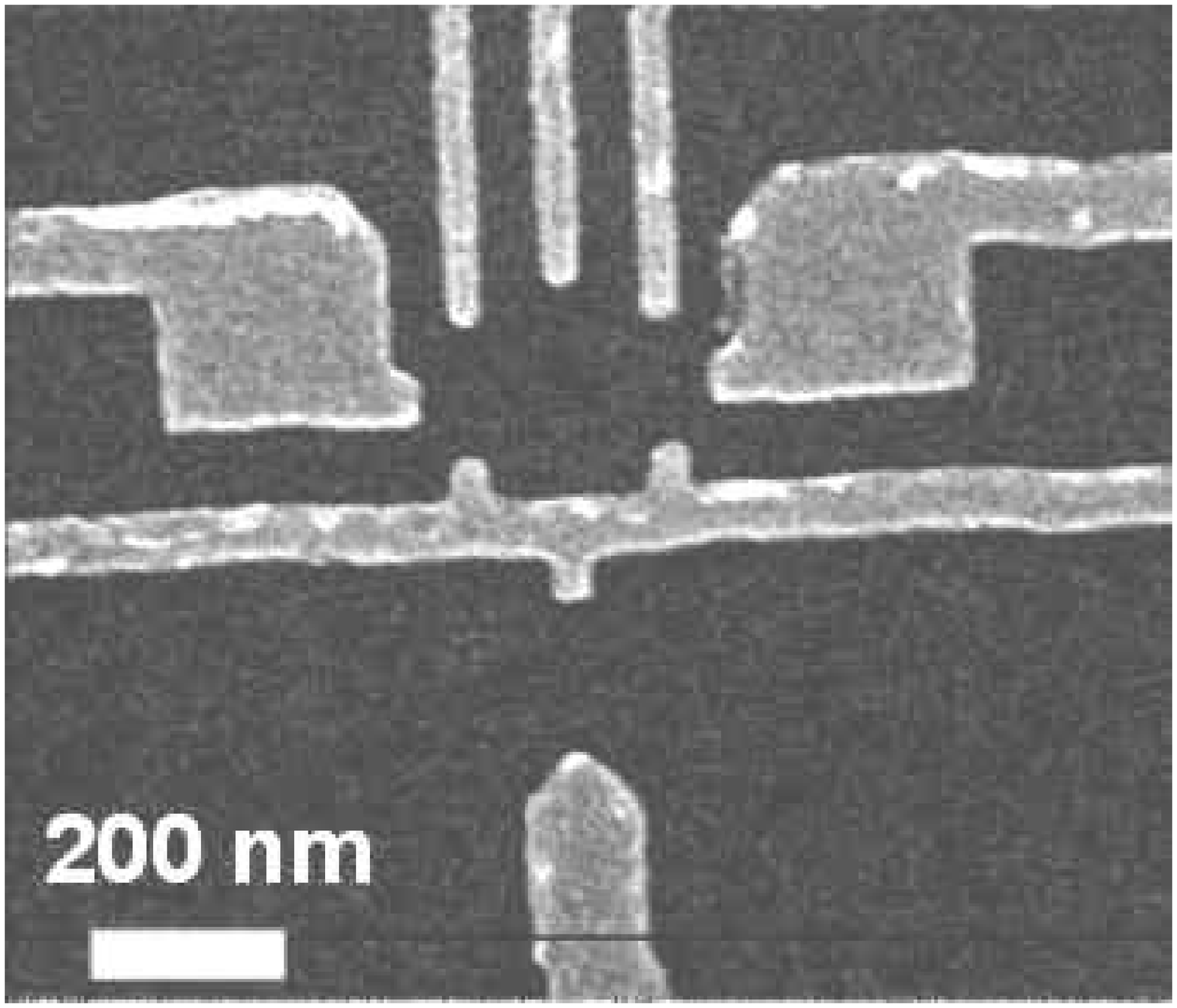}
\hspace{.5cm}
\raisebox{3cm}{(b)}
\includegraphics[width=4cm]{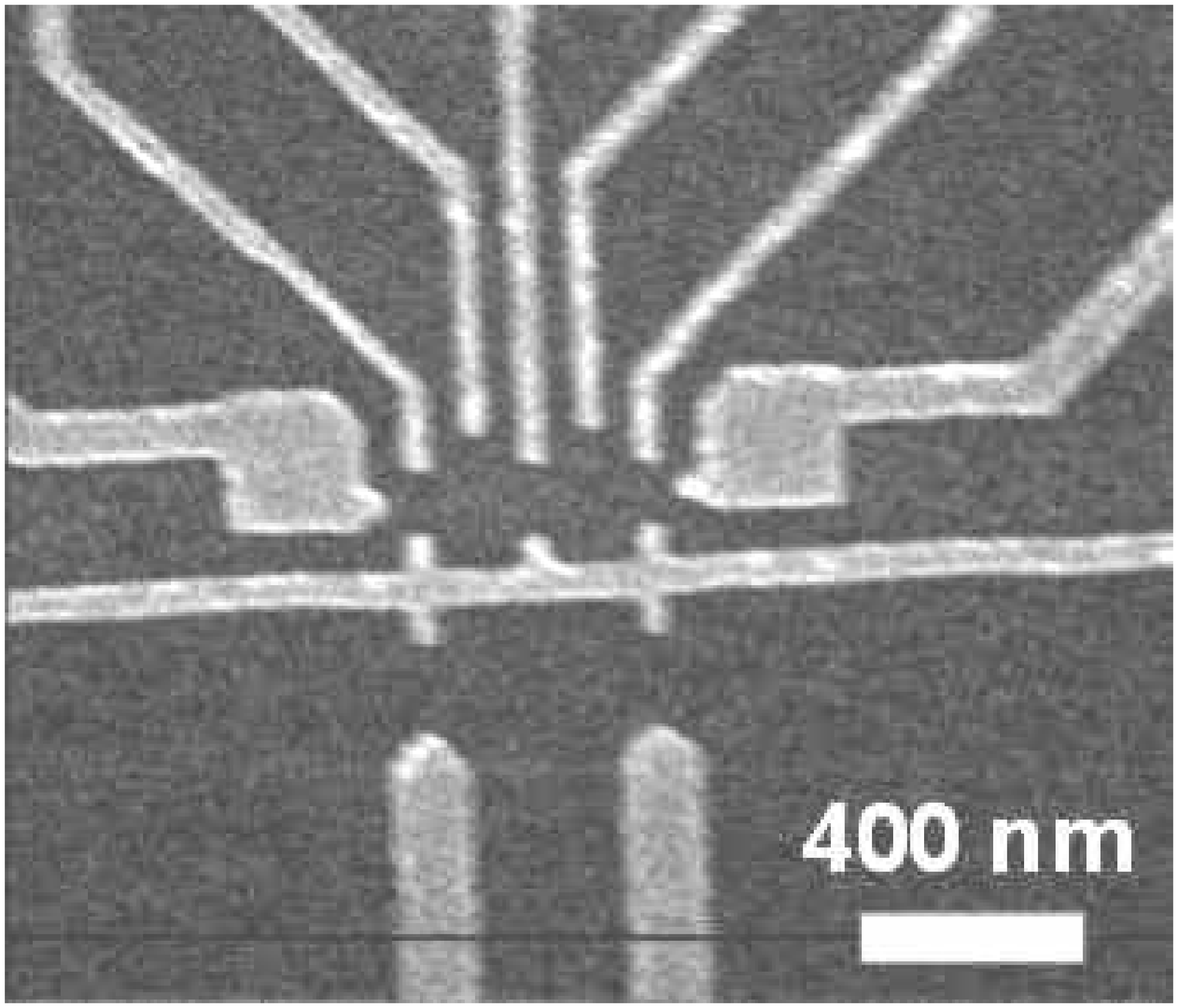}
\end{center}
\caption{SEM images of (a) a single quantum dot and (b) two coupled quantum dots. A quantum point contact placed opposite the quantum dots permits detection of the number of electrons on the dots. The large square gates serve to locally control the electron density in the leads right next to the tunnel barriers to the dots.}
\label{fig:sem_images}
\end{figure}

Initialization to a mixed state can be obtained by letting an electron tunnel to an empty dot from unpolarized or partially polarized leads. The probabilities for $\ua$ and $\da$ electrons to tunnel to the dot depend on the spin-polarization in the leads, on the Fermi level in the leads $E_F$ relative to the $\ket{\ua}$ and $\ket{\da}$ levels in the dot and on the distance between the dot and the $\ket{\ua}$ and $\ket{\da}$ edge states in the leads~\cite{Ciorga00a,McEuen92a,vanderVaart94a} (Fig.~\ref{fig:init_leads} b). These parameters can be tuned via electrostatic gates which control the electron density in the leads and the potential of the dot (Fig.~\ref{fig:sem_images} a).

\section{Read-out}
\label{sec:readout}

The ideal read-out scheme offers high-fidelity, single-shot, strong quantum measurements~\cite{Nielsen00b}. This implies that the measurement must be switched off completely during the computation stage in order to avoid rapid decoherence. Furthermore, the measurement time must be shorter than $T_1$. Finally, the distinction between the $\ket{\ua}$ and $\ket{\da}$ states must be very sharp.

The magnetic moment of a single electron spin is so weak that it has never been detected directly. Previous measurements of single electron spins were based on optical techniques~\cite{Wrachtrup93a,Kohler93a,Gammon01a}. Also, various proposals exist for conversion of spin information to electrical charge~\cite{Kane98a} or current~\cite{Golovach02a}, both of which can be measured with high sensitivity. Here, we propose to use spin-to-charge conversion: the electron on the dot quickly leaves the dot if it is, say, in $\ket{\ua}$, whereas it stays on the dot if it is in $\ket{\da}$. Next, the number of charges on the dot is measured. If there still is one electron charge on the dot, we know the qubit was in $\ket{\da}$, and if there is no charge left, the qubit was in $\ket{\ua}$.

Spin-selective tunneling could in principle be realized using tunnel barriers made from magnetic materials~\cite{Prinz98a}. However, such materials cannot yet be integrated with GaAs/Al$_x$Ga$_{1-x}$As heterostructures. Therefore, we propose instead to operate with spin-polarized leads (quantum Hall regime with $\nu=1$), which provide a reference spin orientation against which the qubit spin orientation can be compared.

In order to motivate our proposed scheme, let us start with some approaches which will {\em not} work. First, if both the $\ket{\ua}$ and $\ket{\da}$ levels in the dot lie below $E_F$, the electron can never escape from the dot. Second, if $E_F$ lies above the $\ket{\ua}$ but below the $\ket{\da}$ dot level, the qubit electron can only tunnel out if it is in $\ket{\da}$; however, as soon as the qubit electron leaves the dot, another electron will enter the dot and occupy $\ket{\ua}$, so we always end up with one electron charge on the dot and no information about the spin state. Third, if $E_F$ lies below the $\ket{\ua}$ and $\ket{\da}$ levels of the dot and $g_{l,\mathrm{eff}} = g_d$, the tunnel process out of the dot is not spin-selective.

\begin{figure}
\begin{center}
\raisebox{2cm}{(a)}
\includegraphics[width=5cm]{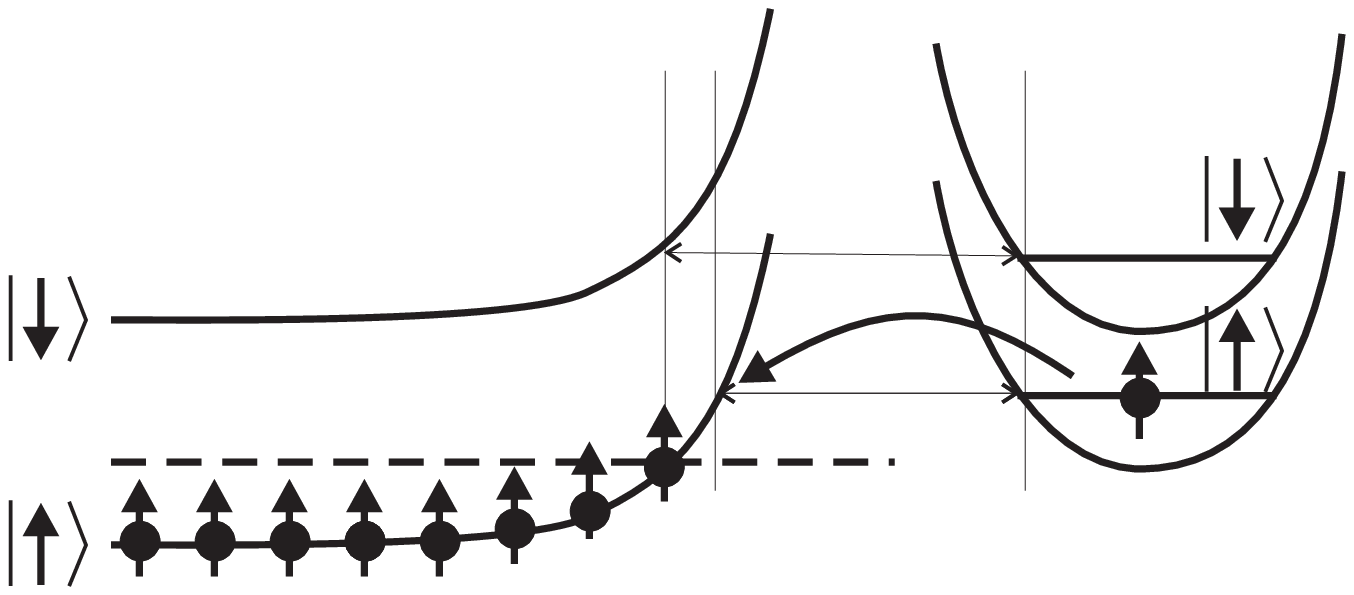}
\hspace{.5cm}
\raisebox{2cm}{(b)}
\includegraphics[width=5cm]{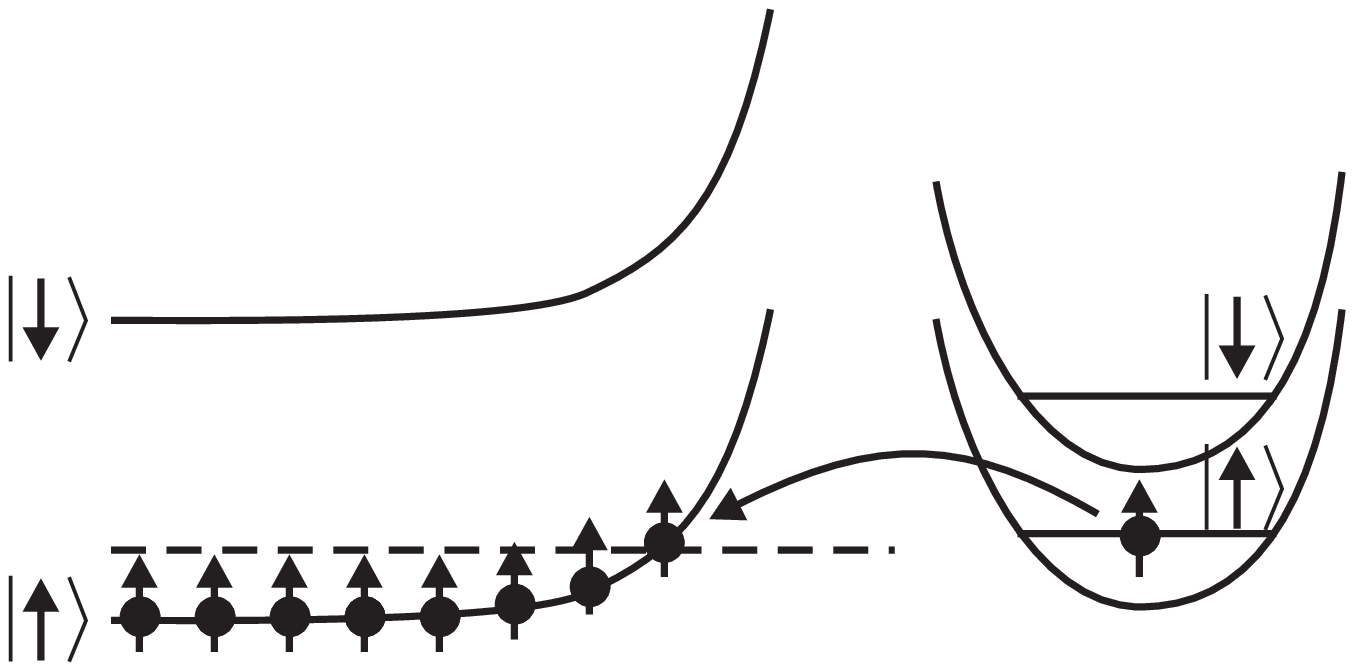}
\end{center}
\vspace*{-2ex}
\caption{(a) When $g_{l,\mathrm{eff}} > g_d$, the dot-lead tunnel rate can be significantly lower for $\ket{\da}$ electrons than for $\ket{\ua}$ electrons, because of the difference in tunnel distances. (b) When $g_{l,\mathrm{eff}} \gg g_d$, the $\ket{\da}$ states in the leads are unreachable energetically for an electron in the dot.}
\label{fig:spin_to_charge}
\end{figure}

However, if $g_{l,\mathrm{eff}} \neq g_d$, the tunnel process {\em is} spin-selective (Fig.~\ref{fig:spin_to_charge}). We recall from section~\ref{sec:initialization} that in $\nu=1$ leads, the exchange interaction increases the total energy of $\da$ electrons with respect to $\ua$ electrons and thus suppresses the $\da$ probability. Similarly, we expect that the increased energy of the $\ket{\da}$ level in the leads reduces the probability that a $\da$ electron will tunnel to the leads from the quantum dot, while $\ua$ electrons can tunnel to the leads easily.

An alternative spin-selective tunneling process, inspired by ~\cite{Golovach02a}, consists of allowing a second electron to tunnel onto the dot from $\nu=1$ leads. This $\ua$ electron will form a singlet state with the qubit electron if the qubit spin was in $\ket{\da}$, whereas they will form a triplet if the qubit was in $\ket{\ua}$. Since the singlet and triplet energies can be made very different~\cite{Tarucha00a}, the second electron will or will not enter the dot depending on the qubit state. As before, subsequent single-shot measurement of the charge on the dot will reveal the qubit state.

We expect both read-out schemes to be very robust: $g_{l,\mathrm{eff}} \mu_B B_0$ is as much as 1 meV (11 Kelvin) for $B_0 = 5$ T and $g_{l,\mathrm{eff}}=5$~\cite{Englert82a}. The singlet-triplet energy difference can also be of order a meV~\cite{Burkard99a,Tarucha00a}. Read-out schemes based on the Zeeman splitting ($\approx 0.1$ meV) are likely less reliable.\\

Measurement of the number of electron charges on the dot must be completed in a time $T_m$ short compared to $T_1$. Furthermore, if the electron is supposed to tunnel, it should do so in a time $T_t$ short compared to $T_m$. If the electron is not supposed to tunnel, it should not do so for a time $T_{nt}$ long compared to $T_m$. In summary, we need $T_t < T_m < T_1, T_{nt}$. Taking $T_1 \ge 100 \mu$s~\cite{Fujisawa02a} and $T_t \approx 0.1 \mu$s, we must measure the charge in a time $T_m$ of $1-10 \mu$s.

A variety of highly sensitive charge meters exist, such as single-electron transistors (SETs), RF-SETs and quantum point contacts (QPCs)~\cite{Sprinzak01a,Field93a}. We have opted to use a QPC (Fig.~\ref{fig:sem_images}), since it is easy to fabricate and integrate with lateral quantum dots, and because we expect its charge sensitivity to be sufficient to determine the number of electron charges on the dot within 1-10 $\mu$s (using a low temperature follower stage with output impedance of a few k$\Omega$).

Finally, we point out that the qubit measurement can be switched off by going to the Coulomb blockade regime with the number of electrons on the dot fixed to one. At this point, charge read-out contains no information about the spin state, and does not cause back-action on the qubit. The qubit measurement is switched on by setting the dot potential so that spin-selective tunneling becomes possible, and spin information is converted into charge information.


\section{Electron spin resonance}

A microwave magnetic field $\vec{B_1}$ oscillating in the plane $\perp$ to $\vec{B_0}$ at a frequency $f_0 = g_d \mu_B B_0 / h$, in resonance with the spin precession about $\vec{B_0}$, will cause the spin to make transitions between $\ket{\ua}$ and $\ket{\da}$. Evidently, if $g_l=g_d$, the spins in the leads would also be on resonance, and the spin polarization in the leads would disappear. Fortunately, $g$ varies with $B_0$, $n$ and the geometry of the structure~\cite{Dobers88a}, thus we expect $g_l \neq g_d$. The choice of $B_0$ strength is a trade-off between reliable initialization and read-out (strong $B_0$ is better) and experimental convenience (low $f_0$ is easier). A field of 5 Tesla seems reasonable: with $g_d=0.44$, high-fidelity initialization and read-out should be within reach (Sections~\ref{sec:initialization} and~\ref{sec:readout}), and $f_0 \approx 30$ GHz.

Properly timed bursts of microwave power tip the spin state over a controlled angle, e.g. $90^\circ$ or $180^\circ$. In order to observe Rabi oscillations, the Rabi period must be at most of the order of the phase randomization time constant $T_2$. The $T_2$ of a single electron spin in a quantum dot has never been measured, but based on experiments with electrons in 2DEG's, we expect that $T_2$ may be $> 100$ ns~\cite{Kikkawa99a}.  For a Rabi period of 150 ns ($f_1 = 6.66$ MHz), we need a microwave field strength $B_1$ of $\approx 1$ mT. For single-qubit rotations much faster than $T_2$, a much stronger $B_1$ is needed.

In order to just observe electron spin resonance (ESR), however, $B_1$ can be much smaller. Steady-state solution of the Bloch equations with a continuous wave (CW) microwave field $B_1$ applied on-resonance with the spin transition, gives Pr$[\ket{\ua}] = [1+1/(1+(2\pi f_1)^2 T_1 T_2)]/2$~\cite{Abragam61a}. Thus, in order to disturb Pr$[\ket{\ua}]$ an observable amount away from $1$, we need $f_1 \ge 1/(2\pi\sqrt{T_1 T_2})$.
Taking $T_1 = 100 \mu$s and $T_2 = 100$ ns, we need $f_1\ge 50$ kHz, and thus $B_1 = 0.01$ mT, one hundred times less than needed to observe Rabi oscillations.\\

Excitation of ESR microwave magnetic fields commonly relies on microwave cavities, but unfortunately, a lot of power is dissipated in metallic cavities: over 1 Watt for $B_1 = 1$ mT at $f_0=30$ GHz and still about 100 $\mu$W for $B_1 = 0.01 $mT~\cite{Jackson98a}. Superconducting cavities are not an option since $B_0$ is too large. 

\begin{figure}
\begin{center}
\raisebox{2cm}{(a)} 
\hspace{1mm}
\raisebox{3mm}{\includegraphics*[height=1.7cm]{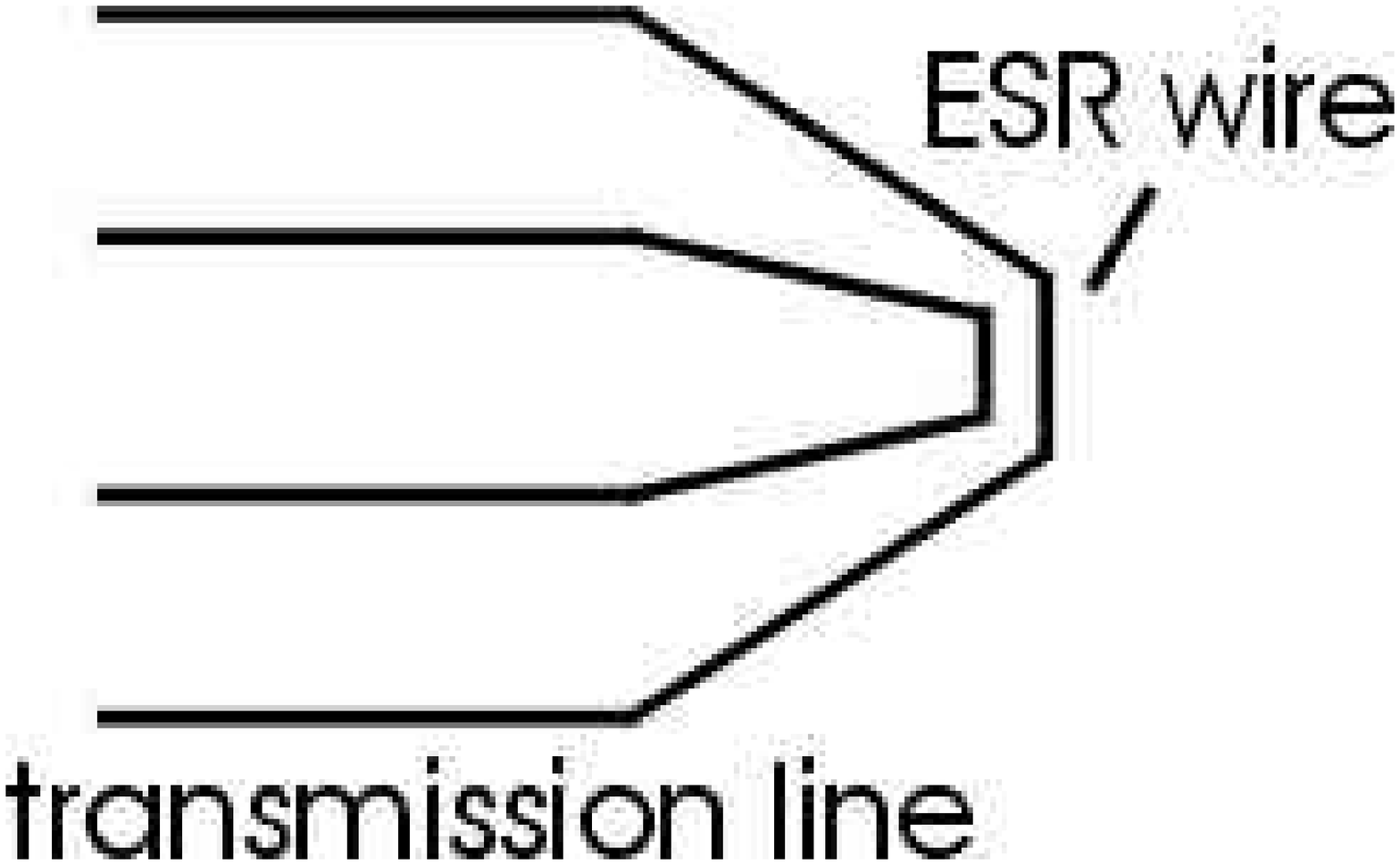}}
\hspace{1cm}
\raisebox{2cm}{(b)}
\hspace{3mm}
\includegraphics*[height=2.2cm]{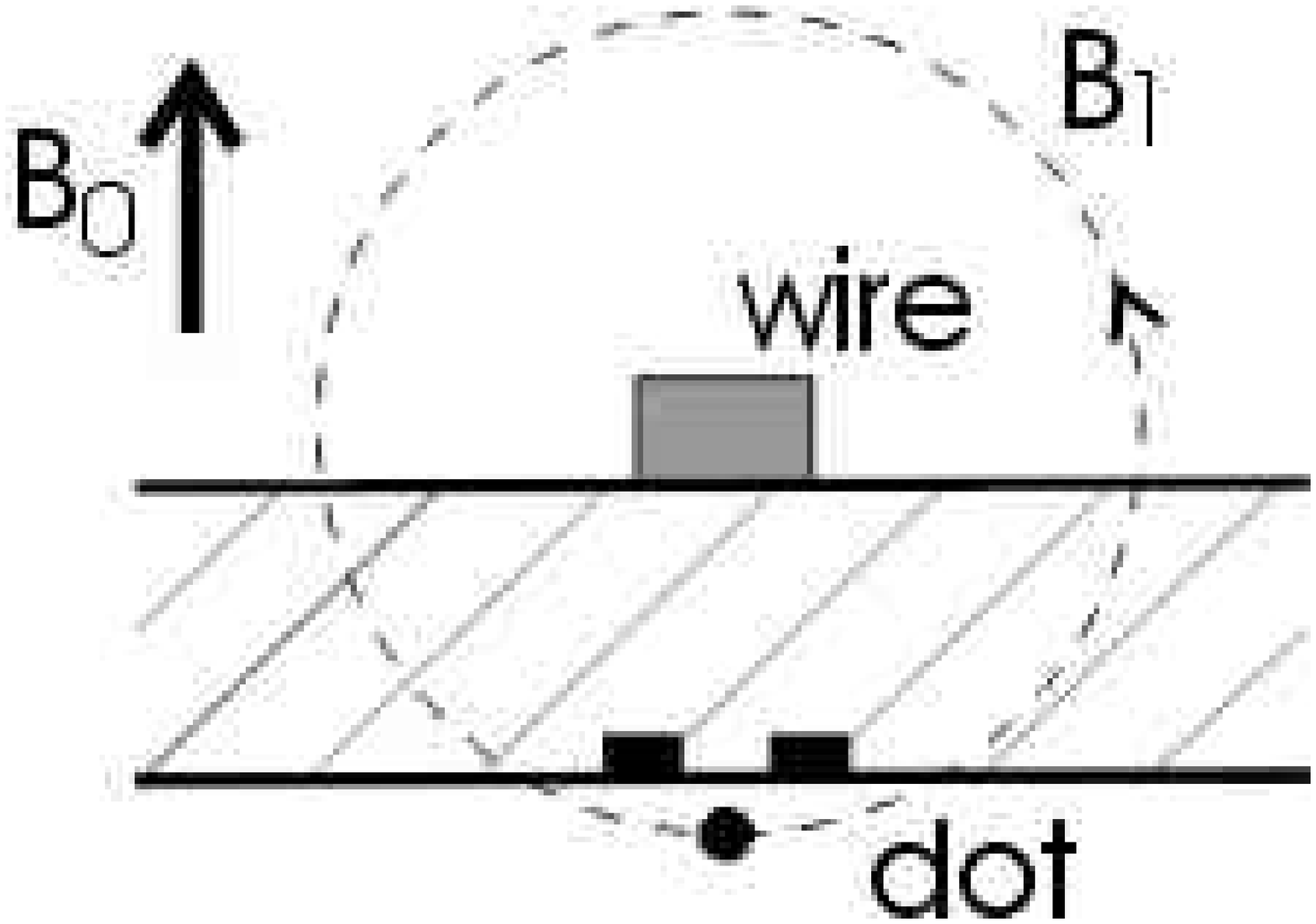}
\end{center}
\vspace*{-0.5cm}
\caption{Schematic drawing of a microfabricated wire deposited on an insulating layer on top of the substrate. (a) Top view and (b) cross-section view.}
\label{fig:wire}
\end{figure}

The alternative we pursue is to send an alternating current through a wire running close by the dot. The wire can be seen as a lumped element if it is much shorter than the wavelength, which is a few mm at 30 GHz near the surface of a GaAs substrate. If the wire terminates a $50 \Omega$ transmission line and has an impedance $\ll 50 \Omega$, it represents a shorted termination and the current is maximum at the wire (Fig.~\ref{fig:wire} a). 

If the wire is placed well within one wavelength $\lambda$ from the quantum dot, the dot is in the near-field region and the electric and magnetic field distribution produced by the AC current is the same as for a DC current~\cite{Jackson98a}. At a distance $r \ll \lambda$ from the center of the wire, a current $I$ through a wire with circular cross-section thus produces a magnetic field $B_1 = \mu I / 2\pi r$, and no electric field. If the wire is located off-chip, say, $200 \mu$m away from the dot, and $\mu$ equals the magnetic susceptibility in vacuum $\mu_0$, we need a current of about 1 A to produce a 1 mT microwave field, and 10 mA to get 0.01 mT. A wire integrated on-chip could be placed much closer to the dot, say 200 nm away, so the current needed is only $\approx 1$ mA or 10 $\mu$A respectively.

Energy dissipation by high frequency currents in transmission lines takes place through ohmic, dielectric and radiation losses~\cite{Jackson98a}. For the on-chip section of the transmission line (a coplanar waveguide or coplanar striplines) and for the wire termination, radiation losses can be kept low by placing the sample inside a cavity with lowest resonance frequency above the operating frequency (longest dimension below 5 mm). At 30 GHz, dielectric losses are typically of the same order as ohmic losses, but in the on-chip section, the wire cross-section is so small that ohmic losses dominate. The off-chip section of the transmission line is a semirigid coaxial connection, which has relatively low loss and is thermally anchored to the higher temperature stages of the cryostat. 

The ohmic losses in the on-chip lines thus dominate; we estimate them to be of order 10 $\mu$W for $I = 1$ mA. Even considering additional power dissipation due to dielectric and other losses, the total dissipated power should be well below the thermal budget at the mixing chamber of a dilution refrigerator (a few 100 $\mu$W at 100 mK), especially if the duty cycle is kept low. Nevertheless, local and instantaneous heating of the device is still a concern so we will keep losses down by maximizing the cross-section of the metal lines, by minimizing $r$, and by constructing a good heat-link from the chip to the mixing chamber.

For $\vec{B_0} \perp$ to the surface (needed for operation in the quantum Hall regime), $\vec{B_1}$ must run through the dot in a direction parallel to the surface, so we must place the wire above the dot, and not to its side. It can be located on top of an insulating dielectric layer (Fig.~\ref{fig:wire} b).

Finally, individual addressing of one among several qubits could be achieved, in principle, via local gate electrodes, which push the individual electron wavefunctions towards a semiconductor layer with a different composition and hence a different $g$-factor (in Al$_x$Ga$_{1-x}$As, $g$ changes from $-0.44$ to $0.4$ when $x$ varies from 0 to 1). In this way, the electrons can be shifted into or away from resonance with the applied microwave field as desired~\cite{Salis01a,Jiang01a}.


\section{The $\sqrt{\mbox{\sc swap}}$ gate}

Two electron spins $S_1$ and $S_2$ in neighbouring quantum dots are coupled to each other by the exchange interaction, which takes the form $J(t) S_1  \cdot S_2$. The result of this interaction is that the states of the two spins are swapped after a certain time. An interaction active for half this time performs the $\sqrt{\mbox{\sc swap}}$ gate, which has been shown to be universal for quantum computation when combined with single qubit rotations~\cite{Burkard99a}. In fact, the exchange interaction is even universal by itself when the state of each qubit is encoded in the state of three electron spins~\cite{DiVincenzo00b}.

The strength $J(t)$ of the exchange interaction depends on the overlap of the respective electron wavefunctions, which varies exponentially with the voltage applied to the gate controlling the tunnel barrier between the two dots. For very small dots with interdot distance of a few tens of nm, $J$ may be as large as 100 GHz~\cite{Burkard99a}. For current sizes of dots and an interdot distance of $\approx 200$ nm, $J$ may still correspond to a frequency of a few tens of GHz, so two-qubit gates could be performed in $< 100$ ps.

We can explore the operation of the {\sc swap} gate as soon as initialization and read-out are reliable, without requiring ESR~\cite{Loss98a}. Imagine qubit 1 is prepared in a pure state $\ket{\ua}$ and qubit 2 is prepared in a statistical mixture of $\ket{\ua}$ and $\ket{\da}$. Measurement of qubit 1 should then always give $\ket{\ua}$ while measurement of qubit 2 should give probabilistically $\ket{\ua}$ or $\ket{\da}$. After application of the {\sc swap} gate, in contrast, measurement of qubit 2 should always give $\ket{\ua}$, while measurement of qubit 1 should give a probabilistic outcome.


\section{Outlook}

In summary, we have presented a procedure for initialization with tunable spin-polarization, for reliable single-shot spin-to-charge conversion and charge measurement, for electron spin resonance with a microfabricated wire and for two-qubit gates using the exchange interaction. All these schemes can be realized with current technology and we have begun experiments using the devices shown in Fig.~\ref{fig:sem_images}. 

The estimated quantum gate duration achievable at present is a few tens of ns. This requires $T_2 > 100 \mu$s in order to reach error probabilities per operation below $10^{-4}$, needed for fault-tolerant quantum computation~\cite{Nielsen00b}. Speeding up single-spin rotations requires much stronger microwave fields, which probably means that alternative ways for excitation must be found which dissipate less power than cavities or wires.

The long term promise of electron spin qubits in GaAs/AlGaAs quantum dots clearly depends crucially on the coherence times $T_2$ and $T_1$. Theoretical studies of decoherence of electron spins in quantum dots have focussed mostly on energy exchange with the bath, i.e. $T_1$~\cite{Khaetskii00a}. The decay time for an electron in an excited orbital state to the ground orbital state has been measured to exceed 200 $\mu$s when a spin flip was involved in the transition~\cite{Fujisawa02a}, so this gives hope that the $T_1$ (spin relaxation time for an electron in the Zeeman split orbital ground state) may also be $> 100 \mu$s. The $T_2$ for isolated electron spins has never been measured --- it is one of the primary objectives of our work.

\begin{acknowledgments}
We thank C.J.P.M. Harmans, W.G. van der Wiel, M. Blaauboer, S. Tarucha, and D. Loss for useful discussions. We acknowledge financial support from the DARPA grant DAAD19-01-1-0659 of the QuIST program, the Dutch Organization for Fundamental Research on Matter (FOM), and the European Union through a TMR Program Network.
\end{acknowledgments}

\bibliographystyle{unsrt}

\begin{thebibliography}{10}

\bibitem{Nielsen00b}
M.A. Nielsen and I.L. Chuang.
\newblock {\em Quantum computation and quantum information}.
\newblock Cambridge University Press, Cambridge, England, 2000.

\bibitem{Loss98a}
D. Loss and D.P. DiVincenzo.
\newblock Quantum computation with quantum dots.
\newblock {\em Phys. Rev. A}, 57:120--126, 1998.

\bibitem{Burkard99a}
G.~Burkard, D.~Loss, and D.P. DiVincenzo.
\newblock Coupled quantum dots as quantum gates.
\newblock {\em Phys. Rev. B}, 59:2070--2078, 1999.

\bibitem{Golovach02a}
V.N. Golovach and D.~Loss.
\newblock Electron spins in artificial atoms and molecules for quantum
  computing.
\newblock {\em Semicond. Sci. Tech.}, 17:355--366, 2002.

\bibitem{Kouwenhoven97a}
L.P. Kouwenhoven {\it et al}.
\newblock Electron transport in quantum dots.
\newblock In L.P.~Kouwenhoven L.L.~Sohn and G.~Sch\"on, eds., {\em
  Proc. of the NATO Advanced Study Institute on Mesoscopic Electron
  Transport}, pp. 105--214. Kluwer Series, 1997.

\bibitem{Kouwenhoven01a}
L.P. Kouwenhoven, D.G. Austing, and S.~Tarucha.
\newblock Few-electron quantum dots.
\newblock {\em Reports on Progress in Physics}, 64:701--736, 2001.

\bibitem{Ciorga00a}
M.~Ciorga {\it et al}.
\newblock Addition spectrum of a lateral dot from {Coulomb} and spin-blockade
  spectroscopy.
\newblock {\em Phys. Rev. B}, 61:R16315--8, 2000.

\bibitem{Sprinzak01a}
D.~Sprinzak, Y.~Ji, M.~Heiblum, D.~Mahalu, and H.~Shtrikman.
\newblock Charge distribution in a {Kondo} correlated quantum dot.
\newblock {\em Phys. Rev. Lett.}, 88:176805, 2002.

\bibitem{Elzerman02a}
J.M.~Elzerman {\it et al.,}
\newblock {\em \it in preparation}, 2002.

\bibitem{Kikkawa99a}
J.M. Kikkawa and D.D. Awschalom.
\newblock Lateral drag of spin coherence in gallium arsenide.
\newblock {\em Nature}, 397:139--141, 1999.

\bibitem{Englert82a}
T.~Englert, D.C. Tsui, A.C. Gossard, and C.~Uihlein.
\newblock $g$-factor enhancement in the {2D} electron gas in {GaAs/}{AlGaAs}
  heterojunctions.
\newblock {\em Surface Science}, 113:295--300, 1982.

\bibitem{McEuen92a}
P.L. McEuen {\it et al}.
\newblock Self-consistent addition spectrum of a {Coulomb} island in the
  quantum hall regime.
\newblock {\em Phys. Rev. B}, 45:11419--11422, 1992.

\bibitem{vanderVaart94a}
N.C. van~der Vaart {\it et al}.
\newblock Time-resolved tunneling of single electrons between {Landau} levels
  in a quantum dot.
\newblock {\em Phys. Rev. Lett.}, 73:320--323, 1994.

\bibitem{Wrachtrup93a}
J.~Wrachtrup, C.~Vonborczyskowski, J.~Bernard, M.~Orrit, and R~Brown.
\newblock Optical detection of magnetic resonance in a single molecule.
\newblock {\em Nature}, 363:244--245, 1993.

\bibitem{Kohler93a}
J.~Kohler {\it et al}.
\newblock Magnetic resonance of a single molecular spin.
\newblock {\em Nature}, 363:242--244, 1993.

\bibitem{Gammon01a}
D.~Gammon {\it et al}.
\newblock Electron and nuclear spin interactions in the optical spectra of
  single {GaAs} quantum dots.
\newblock {\em Phys. Rev. Lett.}, 86:5176--5179, 2001.

\bibitem{Kane98a}
B.E. Kane.
\newblock A silicon-based nuclear spin quantum computer.
\newblock {\em Nature}, 393:133--137, 1998.

\bibitem{Prinz98a}
G.A. Prinz.
\newblock Magnetoelectronics.
\newblock {\em Science}, 282:1660--1663, 1998.

\bibitem{Tarucha00a}
S.~Tarucha {\it et al}.
\newblock Direct coulomb and exchange interaction in artificial atoms.
\newblock {\em Phys. Rev. Lett.}, 84:2485--2489, 2000.

\bibitem{Fujisawa02a}
T.~Fujisawa, D.G. Austing, Y.~Tokura, Y.~Hirayama, and S.~Tarucha.
\newblock Allowed and forbidden transitions in artificial atoms.
\newblock {\em to appear in Nature}, 2002.

\bibitem{Field93a}
M.~Field {\it et al}.
\newblock Measurements of {Coulomb} blockade with a non-invasive voltage probe.
\newblock {\em Phys. Rev. Lett.}, 70:1311--1314, 1993.

\bibitem{Dobers88a}
M.~Dobers, K.~v.~Klitzing, and G.~Weiman.
\newblock Electron-spin resonance in the 2D electron gas of
  {GaAs-Al$_x$Ga$_{1-x}$} heterostructures.
\newblock {\em Phys. Rev. B}, 38:5453, 1988.

\bibitem{Abragam61a}
A.~Abragam.
\newblock {\em Principles of Nuclear Magnetism}.
\newblock Clarendon Press, Oxford, 1961.

\bibitem{Jackson98a}
J.D. Jackson.
\newblock {\em Classical electrodynamics}.
\newblock Wiley, New York, 1998.

\bibitem{Salis01a}
G.~Salis {\it et al}.
\newblock Electrical control of spin coherence in semiconductor nanostructures.
\newblock {\em Nature}, 414:619--622, 2001.

\bibitem{Jiang01a}
H.W. Jiang and E.~Yablonovitch.
\newblock Gate-controlled electron spin resonance in {GaAs/Al$_x$Ga$_{1-x}$As}
  heterostructures.
\newblock {\em Phys. Rev. B}, 64:041307, 2001.

\bibitem{DiVincenzo00b}
D.P. DiVincenzo, D.P. Bacon, D.A. Lidar, and K.B. Whaley.
\newblock Universal quantum computation with the exchange interaction.
\newblock {\em Nature}, 408:339, 2000.

\bibitem{Khaetskii00a}
A.V. Khaetskii and Y.V. Nazarov.
\newblock Spin relaxation in semiconductor quantum dots.
\newblock {\em Phys. Rev. B}, 61:12639--12642, 2000.

\end{thebibliography}

\end{document}